\documentstyle[12pt]{article}
\parskip=\bigskipamount
\begin{document}
\title{\bf Bohm's interpretation and maximally entangled states.}
\date{}
\author{}
\maketitle
\vglue -1.8truecm
\centerline{\large Thomas Durt\footnote{TENA, TONA Free University of Brussels, Pleinlaan 2, B-1050
Brussels, Belgium.} and Yves Pierseaux\footnote{Fondation Wiener-Anspach,
University of Oxford, 10 Merton street, OX14JJ Oxford, UK.}}\bigskip\bigskip

\section*{Abstract}Several no-go theorems showed the incompatibility between the locality
assumption and quantum correlations obtained from maximally entangled spin states. We analyze these
no-go theorems in the framework of Bohm's interpretation. The mechanism by which non-local
correlations appear during the results of measurements performed on distant parts of entangled
systems is explicitly put into evidence in terms of Bohmian trajectories. It is shown that
a GHZ like contradiction of the type+1=-1 occurs for well-chosen initial positions of the Bohmian
trajectories and that it is this essential non-classical feature that makes it possible to
violate the locality condition.

keywords: Bohmian dynamics, non-locality 

\section*{Introduction.} Several no-go theorems show that it is impossible to conciliate locality
and quantum mechanics. The essence of these theorems is that it is impossible to simulate the
results of observations carried out on distant parts of an entangled quantum system in terms of a
local common cause mechanism. This impossibility can be expressed through the
violation of an inequality \cite{bell,clauser,pito} or, even ``stronger'', of an equality
\cite{ghz,mermin}. Besides it is well known that the Bohmian interpretation according to which
particles have at any time a well-defined position \cite{bohm2} makes it possible to simulate all
the quantum predictions for massive particles in the non-relativistic regime. Therefore, it is
interesting to understand better how non-locality appears in the Bohmian picture. It is well-known
that for entangled systems the velocity associated to the Bohmian trajectories can be non-locally
influenced \cite{dewdney1,dewdney2}. The goal of our paper is to analyse in detail the nature of the
Bohmian trajectories in the specific situations encountered in the formulation of three well-known
no-go theorems, Bell's theorem \cite{bell}, Mermin's theorem \cite{mermin} and GHZ's theorem
\cite{ghz}, in which respectively two, three and four spin 1/2 particles are assumed to be prepared
in maximally entangled states. The knowledge of the initial positions of these particles is
sufficient in Bohm's formulation in order to predict the whole trajectories and also the results of
local spin measurements performed with Stern-Gerlach devices. The specification of the initial
positions can thus be considered as a common cause mechanism which makes it possible to explain all
the quantum correlations. In accordance with the no-go theorems and with the properties of
the Bohmian dynamics, this common cause mechanism may however not be considered to be local. We
shall show explicitly in the first section of this paper how, for well chosen initial positions, a
GHZ like contradiction occurs for a situation during which Bell's inequalities
get violated. This result confirms Hardy's formulation of Bell's inequalities
\cite{hardy} according to which the violation of Bell's inequalities is proportional to the
probability that a GHZ like contradiction of the type +1=-1 occurs \cite{ghz}. We shall also show
explicitly in the second section of this paper how in the situation considered in the no-go theorems
of Mermin
\cite{mermin} and Greenberger, Horne and Zeilinger \cite{ghz} the GHZ like contradiction occurs for
all the initial positions.
\section{Bell's theorem in a Bohmian description.}
\subsection{Simultaneous Stern-Gerlach measurements performed on a pair of entangled particles.}

In the situation described in Bell's theorem \cite{bell}, two particles are prepared in an entangled
spin state and spin measurements are performed simultaneously on each particle.
 The full wave-function associated to such a state can be put in the form: \begin{displaymath}\big|
\Psi({\bf r}_{L},\ {\bf r}_{R},\ t) \big>\ =\end{displaymath}\begin{displaymath}\   a\psi_{++}({\bf
r}_{L},\,{\bf r}_{R},\ t)\cdot\big| + \big>_{\rm L}
\otimes \big| +\big>_{\rm R} \ +\  b\psi_{+-}({\bf
r}_{L},\,{\bf r}_{R},\ t)\cdot\big| + \big>_{\rm L} \otimes \big| -\big>_{\rm R}
\end{displaymath} \begin{equation} +\ c\psi_{-+}({\bf
r}_{L},\,{\bf r}_{R},\ t)\cdot\big| - \big>_{\rm L} \otimes \big| +\big>_{\rm R} \ +\
d\psi_{--}({\bf
r}_{L},\,{\bf r}_{R},\ t)\cdot\big| -
\big>_{\rm L} \otimes \big| -\big>_{\rm R}   \end{equation} where $a,\,b,\,c,\,d$
are complex constants representative of the spin entanglement, the indices $L$ and
$R$ represent arbitrary spatial reference frames in two distant regions, the Left and Right regions,
${\bf r}_{L/R}$ represent the position vector expressed in these frames, and
$\big| +/- \big>_{\rm L/R}$ represent up/down spin states along the (not necessary
parallel) $Z$-axes of these frames. We shall assume for convenience that
Stern-Gerlach devices are disposed at the same distance from the source, along the
axes of propagation $X_{L/R}$ of the particles, that the point of penetration inside them
coincides with the origins of the spatial frames, and with the origin of time, and
that the initial state consists of a pair of Gaussian shaped particles of mass $m$ propagating at a
speed
$v_0$ : \begin{equation} \Psi({\bf r}_{L},\ {\bf r}_{R},\ t) \big>\ =\  {1\over (2\sqrt{\pi}\,\delta
r_0)^{6/2}}\ exp({-{\bf r}^{2}_{L}-{\bf r}^{2}_{R}\over 4\delta
r_0^2})\,exp(ik_o(x_{L}+x_{R})).\end{equation} When the $Z_{L/R}$ axis is parallel to the $L/R$
magnetic field, it can be shown that the equation of evolution is separable into the coordinates
$(x_L,\,y_L,\,z_L, x_R,\,y_R,\,z_R).$ For reasons of convenience, we shall also assume that
the Stern-Gerlach set-ups are similar in the Left and Right regions.  Then, it is easy to show that
the solution of the equation of evolution given in the appendix 2 can be expressed in terms of the
solution of the single particle case:
\begin{displaymath}\big|
\Psi({\bf r}_{L},\ {\bf r}_{R},\ t) \big>\ =\end{displaymath}\begin{displaymath}\   a\psi_{+\
L}({\bf r}_{L},\ t)\cdot\psi_{+\ R}({\bf r}_{R},\ t)\cdot\big| + \big>_{\rm L}
\otimes \big| +\big>_{\rm R} \ +\  b\psi_{+\ L}({\bf r}_{L},\ t)\cdot\psi_{-\
R}({\bf r}_{R},\ t)\cdot\big| + \big>_{\rm L} \otimes \big| -\big>_{\rm R}
\end{displaymath} \begin{equation} +\ c\psi_{-\ L}({\bf r}_{L},\ t)\cdot\psi_{+\
R}({\bf r}_{R},\ t)\cdot\big| - \big>_{\rm L} \otimes \big| +\big>_{\rm R} \ +\
d\psi_{-\ L}({\bf r}_{L},\ t)\cdot\psi_{-\ R}({\bf r}_{R},\ t)\cdot\big| -
\big>_{\rm L} \otimes \big| -\big>_{\rm R},   \end{equation}
with: 
\begin{equation}\psi_{+/-\ L/R}({\bf r}_{L/R},\ t)\,=\, \psi^x_{+/-\ L/R}(x_{L/R})\cdot\psi^y_{+/-\
L/R}(y_{L/R},\ t)\cdot \psi^z_{+/-\ L/R}(z_{L/R})\end{equation} where $\psi_{+/-}^x,\ \psi_{+/-}^y,$
and
$\psi_{+/-}^z$ are the solutions of the single particle case that were originally derived
by Bohm \cite{bohm1}. This derivation is reproduced in detail in the appendix 1.

\subsection{Bohm's interpretation.}

Bohm's interpretation \cite{bohm2} is, in summary, the following: whenever we can
associate to the equation of evolution of a quantum system a conservation
equation of the form $\partial_t\,\rho\ =\ div({\bf {\cal J}})$, where $\rho$ is
a positive definite density of probability, and $div({\bf {\cal J}})$ is the
divergence of a current vector, we can interpret this density as a distribution
of localised material points moving with the velocity ${{\bf {\cal J}}\over
\rho}$. It is thus possible to formulate a hidden variable theory for the
system: it would consist of a spatial distribution of material points which 
initially coincides with the quantum distribution (given by $\rho$); these points
move with a velocity equal to ${\bf {\cal J}}/\rho$. In virtue of the
conservation equation, the spatial distribution deduced from this evolution
coincides then for all times with the quantum distribution. According to de Broglie, all
the measurements being, in last resort, position measurements, this hidden
variable theory is, for what concerns practical purposes, equivalent to
orthodox quantum mechanics. 

In the appendix 2, we show how to deduce such a conservation equation for the pair of particles
considered here. From this equation of conservation, it is straightforward to deduce the conserved
density (eqn 50):
 
\begin{displaymath}\rho({\bf
r}_{L},\ {\bf r}_{R},\ t)\ =\ \big<\Psi({\bf r}_{L},\ {\bf r}_{R},\ t)\big|
\Psi({\bf r}_{L},\ {\bf r}_{R},\ t) \big>\
=\end{displaymath}\begin{displaymath}\   |a\psi_{+\ L}({\bf r}_{L},\
t)\cdot\psi_{+\ R}({\bf r}_{R},\ t)|^2\ +\  |b\psi_{+\ L}({\bf r}_{L},\
t)\cdot\psi_{-\ R}({\bf r}_{R},\ t)|^2 \ +\end{displaymath}\begin{equation}
|c\psi_{-\ L}({\bf r}_{L},\ t)\cdot\psi_{+\ R}({\bf r}_{R},\ t)|^2\ +\ |d\psi_{-\
L}({\bf r}_{L},\ t)\cdot\psi_{-\ R}({\bf r}_{R},\ t)|^2 \end{equation} This
density depends in general on the two positions $({\bf r}_{L},\ {\bf r}_{R})$. The
6-dimensional velocity (eqn 51) associated to the conserved density through the relation ${\bf {\cal
J}}\,=\,\rho\cdot\,\vec v$ can be splitted into
two local 3-dimensional velocities $\vec v_L,\ \vec v_R$ defined as follows: 
\begin{displaymath}\vec v_L({\bf r}_{L},\ {\bf r}_{R},\ t)\ = \ {\hbar\over
m\rho} (\Im (\psi_{+\ L}({\bf r}_{L},\ t)^*\ \vec\nabla_L\ \psi_{+\ L}({\bf
r}_{L},\ t))\cdot (|a \psi_{+\ R}({\bf r}_{R},\ t)|^2\ +\ |b \psi_{-\ R}({\bf
r}_{R},\ t)|^2)\end{displaymath}\begin{equation} +\  \Im (\psi_{-\ L}({\bf
r}_{L},\ t)^*\ \vec\nabla_L\ \psi_{-\ L}({\bf r}_{L},\ t))\cdot (|c \psi_{+\
R}({\bf r}_{R},\ t)|^2\ +\ |d \psi_{-\ R}({\bf r}_{R},\ t)|^2)   )\end{equation} 
\begin{displaymath}\vec v_R({\bf r}_{L},\
{\bf r}_{R},\ t)\ = \ {\hbar\over m\rho} (\Im (\psi_{+\
R}({\bf r}_{R},\ t)^*\ \vec\nabla_R\ \psi_{+\ R}({\bf r}_{R},\ t))\cdot (|a
\psi_{+\ L}({\bf r}_{L},\ t)|^2\ +\ |c \psi_{-\ L}({\bf r}_{L},\
t)|^2)\end{displaymath}\begin{equation}\ +\  \Im ( \psi_{-\ R}({\bf r}_{R},\
t)^*\ \vec\nabla_R\ \psi_{-\ R}({\bf r}_{R},\ t))\cdot (|b \psi_{+\ L}({\bf
r}_{L},\ t)|^2\ +\ |d \psi_{-\ L}({\bf r}_{L},\ t)|^2)   \end{equation}where
$\vec\nabla_{L(R)}$ is the gradient on the spatial Left (Right) coordinates, and
$\Im(z) $ the imaginary part of $z$. In general, these velocities contain a
non-vanishing contribution depending on the position of the particle situated at
the other side of the source, and are thus influenced by what happens in a
region from which they are separated by a space-like distance. In the formulation of Bell's theorem,
the particles are assumed to be prepared in the so-called singlet state for which
$a\,=\,d\,=\,{1\over {\sqrt2}}\cdot sin({\theta\over 2})$, and $b\,=\,-c\,=\,{1\over
{\sqrt2}}\cdot cos({\theta\over 2})$ where
$\theta$ is the angle between the magnetic fields in the Left and Right Stern-Gerlach devices. When
these axes are parallel, we recover the standard expression:
\begin{displaymath}\big| \Psi({\bf
r}_{L},\ {\bf r}_{R},\ t) \big>\ =\end{displaymath}\begin{equation}\     {1\over {\sqrt2}}\psi_{+\
L}({\bf r}_{L},\ t)\cdot\psi_{-\ R}({\bf r}_{R},\ t)\cdot\big| + \big>_{\rm L} \otimes \big|
-\big>_{\rm R}
 - \ {1\over {\sqrt2}}\psi_{-\ L}({\bf r}_{L},\ t)\cdot\psi_{+\
R}({\bf r}_{R},\ t)\cdot\big| - \big>_{\rm L} \otimes \big| +\big>_{\rm R}   \end{equation}
It is worth noting that when the axes of the devices are parallel, the results of spin measurements
are always opposite. This perfect correlation is due to the fact that the singlet state is
maximally entangled, and that its expression is invariant under a simultaneous rotation of the
axes of quantization, when both axes are parallel. The specification of $a$, $b$, $c$ and $d$
allows us to express explicitly the Bohmian dynamics during the passage through the Stern-Gerlach
devices. We obtain, after some lenghty but simple computations, the following velocities for the
Left as well as for the Right $x$ and $y$ components:
\begin{equation} {dx\over dt }\ =\ v_0\ +\ {k^2\over 1\ +\ k^2t^2}\,t\ (x\,-\,v_0t)
\end{equation} \begin{equation} {dy\over dt }\ =\ {k^2\over 1\ +\ k^2t^2}\,ty\end{equation}(where
$k\ =\ {\hbar\over 2m\delta^2r_0}$ expresses the diffusion inherent to the Schr\"odinger
equation). The solutions of these dynamical equations are: \begin{equation}x\ =\ v_0 t\ +\ 
x_0 \sqrt{1\,+\,k^2t^2}\end{equation} \begin{equation}y\ =\ y_0 \sqrt
{1\,+\,k^2t^2}\end{equation} In the appendix 1 we show that, during a typical Stern-Gerlach
experiment, the diffusion is negligible during the time of flight through the Stern-Gerlach. The
trajectories along these axes are thus essentially free translations, and present no special
interest for the understanding of the measurement process.

In the singlet state, the dynamics of
the $z$-components is given by: \begin{displaymath} {dz_L\over dt }\ =\
{k^2\over 1\ +\ k^2t^2}\,tz_L\ +\end{displaymath}
\begin{displaymath} {\,+\,s^2 e^{({\beta t^2 \over 1\
+\ k^2t^2} {(z_L\,+\,z_R)\over 2})}\,+\,c^2e^{({\beta t^2 \over 1\
+\ k^2t^2} {(z_L\,-\,z_R)\over 2})}\,-\,c^2 e^{({\beta t^2 \over 1\
+\ k^2t^2} {(-z_L\,+\,z_R)\over 2})} \,-\,s^2 e^{({\beta t^2 \over 1\
+\ k^2t^2} {(-z_L\,-\,z_R)\over 2})}
\over \,+\,s^2 e^{({\beta t^2 \over 1\
+\ k^2t^2} {(z_L\,+\,z_R)\over 2})}\,+\,c^2e^{({\beta t^2 \over 1\
+\ k^2t^2} {(z_L\,-\,z_R)\over 2})}\,+\,c^2 e^{({\beta t^2 \over 1\
+\ k^2t^2} {(-z_L\,+\,z_L)\over 2})} \,+\,s^2 e^{({\beta t^2 \over 1\
+\ k^2t^2} {(-z_R\,-\,z_L)\over 2})}}\end{displaymath}\begin{equation}\cdot\{ {-k^2\over 1\ +\
k^2t^2}\,\alpha t^3\,+\,2\alpha t\}\end{equation}
\begin{displaymath} {dz_R\over dt }\ =\ {k^2\over 1\ +\
k^2t^2}\,tz_R\ +\end{displaymath}
\begin{displaymath} {\,+\,s^2 e^{({\beta t^2 \over 1\
+\ k^2t^2} {(z_L\,+\,z_R)\over 2})}\,-\,c^2e^{({\beta t^2 \over 1\
+\ k^2t^2} {(z_L\,-\,z_R)\over 2})}\,+\,c^2 e^{({\beta t^2 \over 1\
+\ k^2t^2} {(-z_L\,+\,z_L)\over 2})} \,-\,s^2 e^{({\beta t^2 \over 1\
+\ k^2t^2} {(-z_R\,-\,z_L)\over 2})}
\over \,+\,s^2 e^{({\beta t^2 \over 1\
+\ k^2t^2} {(z_L\,+\,z_R)\over 2})}\,+\,c^2e^{({\beta t^2 \over 1\
+\ k^2t^2} {(z_L\,-\,z_R)\over 2})}\,+\,c^2 e^{({\beta t^2 \over 1\
+\ k^2t^2} {(-z_L\,+\,z_R)\over 2})} \,+\,s^2 e^{({\beta t^2 \over 1\
+\ k^2t^2} {(-z_L\,-\,z_R)\over 2})}}\end{displaymath}\begin{equation} \cdot\{ {-k^2\over 1\ +\
k^2t^2}\,\alpha t^3\,+\,2\alpha t\}\end{equation}
\noindent where $\alpha\ =\ a_1\mu/2m$, $\beta\ =\ a_1\mu/m\delta r^2_0$, $c\,=\,cos({\theta\over
2})$ and
$s\,=\,sin({\theta\over
2})$. 

Numerical computations (confirmed by a careful evaluation of
the terms present in the last equations) show that, when the two
magnets act simultaneously, there exist four attractors for the trajectories, corresponding to
the results of spin measurements $(up_L,\ up_R)$, $ (up_L,\ down_R)$, $(down_L,\ up_R)$ and
$(down_L,\ down_R)$. During the passage through the device, the trajectories can be shown to fall
very quickly in the basin of one of these attractors. For instance, in the attractor basin of the
$(up_L,\ up_R)$ outcome, the exponential factor $ e^{({\beta t^2 \over 1\
+\ k^2t^2} {(z_L\,+\,z_R)\over 2})}$ will dominate all the other exponential factors, and
the $z$-velocities will obey the following equations:
\begin{displaymath} {dz_L\over dt }\ =\
{k^2\over 1\ +\ k^2t^2}\,tz_L\ + 1\cdot \{ {-k^2\over 1\ +\
k^2t^2}\,\alpha t^3\,+\,2\alpha t\}\end{displaymath}
\begin{displaymath} {dz_R\over dt }\ =\ {k^2\over 1\ +\
k^2t^2}\,tz_R\ + 1\cdot \{ {-k^2\over 1\ +\
k^2t^2}\,\alpha t^3\,+\,2\alpha t\}\end{displaymath}
In virtue of the fact that, in a well-conceived device, we can neglect the diffusion terms
relatively to the classical drift terms, we can take $k$ to be equal to 0 in the previous equation,
which gives the classical expression for the velocities ($2\alpha$ is the magnitude of the classical
acceleration undergone inside the magnetic field of the device): 
\begin{displaymath} {dz_L\over dt }\ =\ +\,2\alpha t\end{displaymath}
\begin{displaymath} {dz_R\over dt }\ =\ \,+\,2\alpha t\end{displaymath}
Similarly, in say the attractor basin of the $(down_L,\ up_R)$ outcome, we shall get:
\begin{displaymath} {dz_L\over dt }\ =\ -\,2\alpha t\end{displaymath}
\begin{displaymath} {dz_R\over dt }\ =\ \,+\,2\alpha t\end{displaymath}
This means that after a short time, the trajectories become and remain quasi-classical. The choice
of the attractor basin is deterministically speficied by the initial positions. It is worth noting
that when the initial positions lie in between the regions that are characterised by this
quasi-classical behavior the deflection will be quite less important. These positions can be
associated to the tails of the outgoing wave-packets and their weight is negligible, because the
diffusion of the Gaussian wave-packets is negligible relatively to the average deflection undergone
during the passage through the Stern-Gerlach device. 
\subsection{GHZ like contradiction in the two particles case.}
In last resort, the final outcome is thus
``nearly always'' unambiguously determined by the initial positions
$z_L$ and $z_R$. Let us assume that we choose to change the direction of the magnet in the Right
device. Then, the final outcome is determined by the initial positions $z_L$ and $z'_R$. $z'_R$
depends on
$y_R$ and $z_R$ and on the angle between the old quantization axis
$Z_R$ and the new one $Z'_R$. It could happen that for the same initial position
coordinates $z_L,\,y_R,\,z_R$, the attractor basin of the Left trajectory changes due to the change
of
$z_R$ into $z'_R$. This means that the choice of the basis of detection in the Right region can
non-locally influence the outcome of the measurement in the Left region. Therefore although the
results of all possible measurements can be deterministically foreseen on the basis of the
knowledge of the initial positions, which constitutes a common-cause mechanism, this mechanism is
non-local. This helps to understand how Bohmian mechanics makes it possible to violate Bell's
inequalities. The probability of an outcome can be obtained, after integration of the Bohmian
velocities, by performing a weighted average on the predictions associated to this outcome for
particular initial positions. The weight is equal to the initial probability of presence, a
Gaussian distribution in our case. In accordance with Bohm's interpretation, the probability
obtained so is exactly the same as the standard quantum probability. In
\cite{durt}, the attractor basins of the outcomes related to different non-compatible spin
measurements in the Left and Right region were explicitly determined for a well-chosen initial
position and the following result was shown. If we choose four directions ($Z_L,\ Z'_L,\ Z_R,\
Z'_R$) such that the directions
$Z_L,\ Z'_L,\ Z'_R$ are coplanar and are all separated by angles of 120 degrees,
while $Z'_L$ and $Z_R$ are parallel, that the system of two particles is prepared in the singlet
state, and that the initial coordinates of position of the pair are defined as follows:
$(y_L,\ z_L)(t\ =\ 0)\ =\ 10^{-3}cm\cdot ({\sqrt 3}/2,\ 1/2)$, and $(y_R,\ z_R)(t\ =\ 0)\ =\
1.1\cdot10^{-3}cm\cdot(\ {\sqrt 3}/2,\ 1/2)$ then we shall observe, after integration of the
Bohmian velocities, the following outcomes: $(up_{Z_L},\ up_{Z_R}),\ (up_{Z_L'},\ up_{Z_R'})$, and
$(down_{Z_L'},\ up_{Z_R})$. If the locality assumption was valid, the results of the measurements
ought to be predetermined, before the measurements take place. This can be shown by an EPR like
reasoning \cite{epr} which is, roughly summarised, the following: -in the singlet state, the value
of the spin component along an arbitrary axis of reference in the Left region can be deduced from
the observation of the corresponding value in the Right region because the singlet state exhibits
perfect (100
\%) correlations; -if we assume that distant measurements may not influence the local spin values,
then these values existed necessarily before their measurement took place. This means that the local
value (in the Left region but also in the Right region by a similar reasoning) of the spin along an
arbitrary direction ought to be predetermined before the measurement.

Obviously, if the locality assumption was valid, we could deduce from the predictions relative to
the two first joint measurements ($(up_{Z_L},\ up_{Z_R}),\ (up_{Z_L'},\ up_{Z_R'})$), that the third
result ought to be
$(up_{Z_L'},\ up_{Z_R})$ too, which is not the case here (we find $(down_{Z_L'},\ up_{Z_R})$). In
\cite{durt} we showed how Bohmian non-locality is related to the non-factorisability condition
postulated by Bell as a consequence of the locality condition. For the purposes of the present
paper, we went further. We also integrated the velocities when the magnets are parallel to the
direction
$(Z_L,\,Z_R')$, and we found the outcome $(up_{Z_L},\ up_{Z_R'})$. By assigning a value +1 to the
outcome spin up and -1 to the outcome spin down, we arrive to a
GHZ like contradiction of the type +1=-1 that we shall study in detail in the next section. This
can be done by considering the product of the outcomes of
$Z_L\,\cap\,Z_R$ and
$Z'_L\,\cap\,Z'_R$ at one side and the product of the outcomes of $Z'_L\,\cap\,Z_R$ and
$Z_L\,\cap\,Z_R'$ at the other side. If the outcomes associated to the four experiments were
predetermined, these products ought clearly to be equal but for the choice of initial position
considered here they are respectively equal to +1 and -1, which can be put in the paradoxical form
+1=-1. 

It can be shown \cite{durt} that when the four directions ($Z_L,\ Z'_L,\ Z_R,\
Z'_R$) are coplanar and are all separated by angles of 120 degrees,
while $Z'_L$ and $Z_R$ are parallel, and that the system of two particles is prepared in the singlet
state, Bell's and Clauser-Horne's inequalities are violated. It can also be shown that a GHZ like
contradiction occurs only for a fraction of the initial positions that differs significatively
from unity. For instance when the initial coordinates of position of the pair are defined as
follows:
$(y_L,\ z_L)(t\ =\ 0)\ =\ 10^{-3}cm\cdot ({\sqrt 3}/2,\ 1/2)$, and $(y_R,\ z_R)(t\ =\ 0)\ =\
-1\cdot10^{-3}cm\cdot(\ {\sqrt 3}/2,\ 1/2)$ then we shall observe, after integration of the
Bohmian velocities, the following outcomes: $(up_{Z_L},\ down_{Z_R}),\ (up_{Z_L'},\ down_{Z_R'})$,
$(up_{Z_L},\ down_{Z_R'})$ and
$(up_{Z_L'},\ down_{Z_R})$. In this case, no GHZ contradiction occurs, and the results of local
measurements do not depend on which measurements are performed in a distant region. These results
confirm Hardy's analysis who showed
\cite{hardy} that the degree of violation of Bell's and Clauser-Horne's inequalities is
proportional to the probability of encountering GHZ like paradoxical situations.

Besides, Peres \cite{peres} showed that for two particles in the singlet state such a paradoxical
situation also characterised the quantum average values of well-chosen observables. The
interpretation of Peres's paradox in the framework of Bohm's theory was already studied by Dewdney
in \cite{dewdney2} so that we shall not repeat his work in the present paper. Furthermore, it is
impossible to express Peres's paradox in terms of local Stern-Gerlach measurements only because it
involves non-local measurements. Therefore, presently, there is no explicit formulation of this
paradox in terms of Bohmian trajectories. Nevertheless, such a formulation is possible in
principle \cite{dewdney2}. It is worth noting that the implications of Bohm's theory regarding
non-locality were already recognised by Bohm and Bell, and that the elucidation of
certain no-go theorems was qualitatively developed in the past (see for instance the references
\cite{dewdney1,dewdney2} and references therein). Computer simulations of the dynamics inside the
Stern-Gerlach devices (with two particles and parallel magnets only) can be found in
\cite{dewdney2}, as well as a qualitative discussion of their connection with Peres's paradox. The
analytical expression of Bohm's velocities in the general two particle case (with parallel and
non-parallel magnets) was given in
\cite{durt}, as well as the connection between Bohmian non-locality and non-factorisability.
Nevertheless, the
occurence of a GHZ like contradiction in the two particles case as well as the explicit study of
the topology of the attractor basins for the three and four particle cases and their connection to
GHZ like contradictions were
not, to the knowledge of the authors, published elsewhere. This brings us to the next chapter.

\section{Mermin and GHZ no-go theorems in a Bohmian description.}

The two paradoxes that we shall study in this section are of the following form: for well-chosen
local observables and quantum states, the quantum correlations are such that if the results of the
local measurements are fixed by advance and do not depend on the measurements that are performed in
distant regions we come to a contradiction of the type +1=-1. We showed in the previous section
that in the two particles case such a contradiction could occur for well-chosen initial positions.
However, after averaging over the initial positions of the particles, for which the contradiction
occurs sometimes but not always, the paradox is expressed by the violation of an
inequality in accordance with Hardy's analysis \cite{hardy}. The novelty of Mermin's and GHZ
paradoxes
\cite{ghz,mermin} is that the contradiction does not take the form of the violation of an
inequality but well of an equality. As we shall now show, the meaning of this result, interpreted
in terms of Bohm's trajectories is that all the initial positions lead to the same paradoxical
situation, and not only some of them as in the two particles case. 
\subsection{Mermin's no-go theorem: GHZ like contradiction in the three particles case.}
In Mermin's theorem \cite{mermin}, three particles are prepared in a maximally entangled spin state
of the form:
\begin{displaymath}\big| \Psi({\bf
r}_{1},\ {\bf r}_{2},\ {\bf r}_{3},\ t) \big>\
=\end{displaymath}
\begin{displaymath}\     {1\over {\sqrt2}}\psi_{+\ 1}({\bf
r}_{1},\ t)\cdot\psi_{+\ 2}({\bf r}_{2},\ t)\cdot\psi_{+\ 3}({\bf r}_{3},\ t)\cdot\big| +
\big>_{\rm 1}
\otimes \big| +\big>_{\rm 2}\otimes \big| +\big>_{\rm 3}\end{displaymath}
\begin{equation}- \
{1\over {\sqrt2}}\psi_{-\ 1}({\bf r}_{1},\ t)\cdot\psi_{-\ 2}({\bf r}_{2},\ t)\cdot\psi_{-\
3}({\bf r}_{3},\ t)\cdot \big| - \big>_{\rm 1} \otimes \big| -\big>_{\rm 2} \otimes \big|
-\big>_{\rm 3},   \end{equation} The particles are sent
along three coplanar directions separated by an angle of 120 degrees and $\big| +\big>_{\rm i}$
($\big| -\big>_{\rm i}$) represents a spin up (down) state along the direction of propagation of the
$i$th particle. Four ``product'' observables each of which is associated to a simultaneous spin
measurement performed on the three particles are considered: $\Sigma^1_x\cdot \Sigma^2_y\cdot
\Sigma^3_y$,
$\Sigma^1_y\cdot\Sigma^2_x\cdot \Sigma^3_y$,
$\Sigma^1_y\cdot\Sigma^2_y\cdot\Sigma^3_x$ and $\Sigma^1_x\cdot\Sigma^2_x\cdot\Sigma^3_x$, where
$\Sigma^i_x$ represents a Stern-Gerlach mesurement performed on the $i$th particle with the magnet
parallel to the
$X$ direction which is orthogonal to the plane of propagation of the particles and $\Sigma^i_y$
represents a Stern-Gerlach mesurement performed on the $i$th particle with the magnet parallel to
the
$Y_i$ direction which is orthogonal to the direction of propagation of the $i$th particle and to
$X$. The $\Sigma$ matrices possess two eigenvalues, +1 and -1. The eigenvalue +1 corresponds to the
outcome spin up and -1 to the outcome spin down. 
$\big|
\Psi({\bf r}_{1},\ {\bf r}_{2},\ {\bf r}_{3},\ t)
\big>$ is simultaneous eigenstate of the four product observables for the eigenvalues $+1,\ +1,\
+1,\ -1$. This imposes severe constraints on the results of local spin measurements. For instance,
without performing the measurement of $\Sigma^1_x$, we can deduce what would be the outcome observed
during this measurement from the results of the measurements of $\Sigma^2_x$ and $\Sigma^3_x$. By
an EPR like reasoning similar to the one described in the previous section, the locality assumption
implies that the outcome of the measurement of $\Sigma^1_x$ must be determined before the
measurement took place and does not depend on which measurement is performed in distant regions on
the other particles.  A similar reasoning holds for the measurement of $\Sigma_y^1$, $\Sigma^2_x$,
$\Sigma^2_y$, $\Sigma^3_x$ and $\Sigma^3_y$. Let us denote
$S_x^i$ and
$S_y^i$ the prediction assigned to the measurement of $\Sigma_x^i$ and $\Sigma_y^i$; the
prediction +1 corresponds to an ``up'' deflection inside the Stern-Gerlach device, and -1 to a
``down'' deflection. The constraints imposed by the properties of the maximally entangled state in
which the particles are prepared can be expressed as follows:
\begin{displaymath}S_x^1\cdot S_y^2\cdot S_y^3\ =\ +1\end{displaymath}
\begin{displaymath}S_y^1\cdot S_x^2\cdot S_y^3\ =\ +1\end{displaymath}
\begin{displaymath}S_y^1\cdot S_y^2\cdot S_x^3\ =\
+1\end{displaymath}
\begin{equation}S_x^1\cdot S_x^2\cdot S_x^3\ =\ -1.\end{equation}
The product of the three first equations gives:

\begin{equation}(S_y^1)^2\cdot(S_y^2)^2\cdot(S_y^3)^2\cdot S_x^1\cdot S_x^2\cdot
S_x^3\ =\ +1\end{equation}
$S_y^i$ being equal to + or - 1, we get that \begin{equation}S_x^1\cdot S_x^2\cdot S_x^3\ =\
+1,\end{equation}which together with the fourth equation leads to a contradiction of the
type +1=-1. We encountered already such a contradiction in the previous section, and showed that
it could be explained in terms of the non-local properties of Bohmian trajectories. This is also
true in the present case. By a straightforward generalisation of the two particles case, it can be
shown that the Bohmian velocities during the measurement of one spin component along
$X$, one component along $Y$ and another one along $Y'$ are the following:
\begin{displaymath} {dx\over dt }\ =\
{k^2\over 1\ +\ k^2t^2}\,tx\ +\end{displaymath}
\begin{displaymath} { \,+\,e^{({\beta t^2 \over 1\
+\ k^2t^2} {(x\,+\,y\,+\,y')\over 2})}\,+\,e^{({\beta t^2 \over 1\
+\ k^2t^2} {(x\,-\,y\,-\,y')\over 2})}\,-\,e^{({\beta t^2 \over 1\
+\ k^2t^2} {(-x\,-\,y\,+\,y')\over 2})}\,-\,e^{({\beta t^2 \over 1\
+\ k^2t^2} {(-x\,+\,y\,-\,y')\over 2})} 
\over \,+\,e^{({\beta t^2 \over 1\
+\ k^2t^2} {(x\,+\,y\,+\,y')\over 2})}\,+\,\,+\,e^{({\beta t^2 \over 1\
+\ k^2t^2} {(x\,-\,y\,-\,y')\over 2})}e^{({\beta t^2 \over 1\
+\ k^2t^2} {(-x\,-\,y\,+\,y')\over 2})}\,+\,e^{({\beta t^2 \over 1\
+\ k^2t^2} {(-x\,+\,y\,-\,y')\over 2})}  } \end{displaymath}
\begin{equation}\cdot\{ {-k^2\over 1\ +\
k^2t^2}\,\alpha t^3\,+\,2\alpha t\}\end{equation}
\begin{displaymath} {dy\over dt }\ =\
{k^2\over 1\ +\ k^2t^2}\,ty\ +\end{displaymath}
\begin{displaymath} { \,+\,e^{({\beta t^2 \over 1\
+\ k^2t^2} {(x\,+\,y\,+\,y')\over 2})}\,-\,e^{({\beta t^2 \over 1\
+\ k^2t^2} {(x\,-\,y\,-\,y')\over 2})}\,-\,e^{({\beta t^2 \over 1\
+\ k^2t^2} {(-x\,-\,y\,+\,y')\over 2})}\,+\,e^{({\beta t^2 \over 1\
+\ k^2t^2} {(-x\,+\,y\,-\,y')\over 2})} 
\over \,+\,e^{({\beta t^2 \over 1\
+\ k^2t^2} {(x\,+\,y\,+\,y')\over 2})}\,+\,e^{({\beta t^2 \over 1\
+\ k^2t^2} {(x\,-\,y\,-\,y')\over 2})}\,+\,e^{({\beta t^2 \over 1\
+\ k^2t^2} {(-x\,-\,y\,+\,y')\over 2})}\,+\,e^{({\beta t^2 \over 1\
+\ k^2t^2} {(-x\,+\,y\,-\,y')\over 2})} }\end{displaymath}\begin{equation}\cdot\{ {-k^2\over 1\ +\
k^2t^2}\,\alpha t^3\,+\,2\alpha t\}\end{equation}
\begin{displaymath} {dy'\over dt }\ =\
{k^2\over 1\ +\ k^2t^2}\,ty'\ +\end{displaymath}
\begin{displaymath} { \,+\,e^{({\beta t^2 \over 1\
+\ k^2t^2} {(x\,+\,y\,+\,y')\over 2})}\,-\,e^{({\beta t^2 \over 1\
+\ k^2t^2} {(x\,-\,y\,-\,y')\over 2})}\,+\,e^{({\beta t^2 \over 1\
+\ k^2t^2} {(-x\,-\,y\,+\,y')\over 2})}\,-\,e^{({\beta t^2 \over 1\
+\ k^2t^2} {(-x\,+\,y\,-\,y')\over 2})} 
\over \,+\,e^{({\beta t^2 \over 1\
+\ k^2t^2} {(x\,+\,y\,+\,y')\over 2})}\,+\,e^{({\beta t^2 \over 1\
+\ k^2t^2} {(x\,-\,y\,-\,y')\over 2})}\,+\,e^{({\beta t^2 \over 1\
+\ k^2t^2} {(-x\,-\,y\,+\,y')\over 2})}\,+\,e^{({\beta t^2 \over 1\
+\ k^2t^2} {(-x\,+\,y\,-\,y')\over 2})} }\end{displaymath}\begin{equation}\cdot\{ {-k^2\over 1\ +\
k^2t^2}\,\alpha t^3\,+\,2\alpha t\}\end{equation}
They possess four attractor basins that correspond to the outcomes (+ + +), (+ - -) (- - +) and
(- + -). During the measurement of the fourth observable $\Sigma^1_x\cdot\Sigma^2_x\cdot\Sigma^3_x$,
the Bohmian trajectories obey: 

\begin{displaymath} {dx_1\over dt }\ =\
{k^2\over 1\ +\ k^2t^2}\,tx_1\ +\end{displaymath}
\begin{displaymath} { \,-\,e^{({\beta t^2 \over 1\
+\ k^2t^2} {(-x_1\,-x_2\,-\,x_3)\over 2})}\,-\,e^{({\beta t^2 \over 1\
+\ k^2t^2} {(-x_1\,+\,x_2\,+\,x_3)\over 2})}\,+\,e^{({\beta t^2 \over 1\
+\ k^2t^2} {(+x_1\,-\,x_2\,+\,x_3)\over 2})}\,+\,e^{({\beta t^2 \over 1\
+\ k^2t^2} {(+x_1\,+\,x_2\,-\,x_3)\over 2})} 
\over \,+\,e^{({\beta t^2 \over 1\
+\ k^2t^2} {(-x_1\,-x_2\,-\,x_3)\over 2})}\,+\,e^{({\beta t^2 \over 1\
+\ k^2t^2} {(-x_1\,+\,x_2\,+\,x_3)\over 2})}\,+\,e^{({\beta t^2 \over 1\
+\ k^2t^2} {(+x_1\,-\,x_2\,+\,x_3)\over 2})}\,+\,e^{({\beta t^2 \over 1\
+\ k^2t^2} {(+x_1\,+\,x_2\,-\,x_3)\over 2})} }\end{displaymath}\begin{equation}\cdot\{ {-k^2\over 1\
+\ k^2t^2}\,\alpha t^3\,+\,2\alpha t\}\end{equation}
\begin{displaymath} {dx_2\over dt }\ =\
{k^2\over 1\ +\ k^2t^2}\,tx_2\ +\end{displaymath}
\begin{displaymath} {\,-\,e^{({\beta t^2 \over 1\
+\ k^2t^2} {(-x_1\,-x_2\,-\,x_3)\over 2})}\,+\,e^{({\beta t^2 \over 1\
+\ k^2t^2} {(-x_1\,+\,x_2\,+\,x_3)\over 2})}\,-\,e^{({\beta t^2 \over 1\
+\ k^2t^2} {(+x_1\,-\,x_2\,+\,x_3)\over 2})}\,+\,e^{({\beta t^2 \over 1\
+\ k^2t^2} {(+x_1\,+\,x_2\,-\,x_3)\over 2})} 
\over \,+\,e^{({\beta t^2 \over 1\
+\ k^2t^2} {(-x_1\,-x_2\,-\,x_3)\over 2})}\,+\,e^{({\beta t^2 \over 1\
+\ k^2t^2} {(-x_1\,+\,x_2\,+\,x_3)\over 2})}\,+\,e^{({\beta t^2 \over 1\
+\ k^2t^2} {(+x_1\,-\,x_2\,+\,x_3)\over 2})}\,+\,e^{({\beta t^2 \over 1\
+\ k^2t^2} {(+x_1\,+\,x_2\,-\,x_3)\over 2})} }\end{displaymath}\begin{equation}\cdot\{ {-k^2\over
1\ +\ k^2t^2}\,\alpha t^3\,+\,2\alpha t\}\end{equation}
\begin{displaymath} {dx_3\over dt }\ =\
{k^2\over 1\ +\ k^2t^2}\,tx_3\ +\end{displaymath}
\begin{displaymath} {\,-\,e^{({\beta t^2 \over 1\
+\ k^2t^2} {(-x_1\,-x_2\,-\,x_3)\over 2})}\,+\,e^{({\beta t^2 \over 1\
+\ k^2t^2} {(-x_1\,+\,x_2\,+\,x_3)\over 2})}\,+\,e^{({\beta t^2 \over 1\
+\ k^2t^2} {(+x_1\,-\,x_2\,+\,x_3)\over 2})}\,-\,e^{({\beta t^2 \over 1\
+\ k^2t^2} {(+x_1\,+\,x_2\,-\,x_3)\over 2})} 
\over \,+\,e^{({\beta t^2 \over 1\
+\ k^2t^2} {(-x_1\,-x_2\,-\,x_3)\over 2})}\,+\,e^{({\beta t^2 \over 1\
+\ k^2t^2} {(-x_1\,+\,x_2\,+\,x_3)\over 2})}\,+\,e^{({\beta t^2 \over 1\
+\ k^2t^2} {(+x_1\,-\,x_2\,+\,x_3)\over 2})}\,+\,e^{({\beta t^2 \over 1\
+\ k^2t^2} {(+x_1\,+\,x_2\,-\,x_3)\over 2})} }\end{displaymath}\begin{equation}\cdot\{ {-k^2\over
1\ +\ k^2t^2}\,\alpha t^3\,+\,2\alpha t\}\end{equation}
They possess four attractor basins that correspond to the outcomes (- - -), (- + +), (+ - +), and
(+ + -). Very quickly, the Bohmian trajectory will fall into one of these basins and follow a
nearly-classical dynamics. For each choice of initial position (excepted for the initial positions
that belong to the regions which separate two basins, which are regions of negligible weight that
correspond to the tails of the Gaussian packets), the product of the predicted outcomes associated
to the observables $\Sigma_x^1$, $\Sigma_x^2$ and $\Sigma_x^3$ is equal to +1 when we consider their
realisation during the three first experiments and to -1 when we consider the last one during which
they are simultaneously measured.
\subsection{Greenberger, Horne and Zeilinger's no-go theorem: GHZ like contradiction in the four
particles case.}

For what concerns the GHZ paradox \cite{ghz}, we shall firstly reformulate it in a simplified form
that is closer to Mermin's formulation. Four particles are now prepared in a maximally entangled
spin state of the form:
\begin{displaymath}\big| \Psi({\bf
r}_{1},\ {\bf r}_{2},\ {\bf r}_{3},\ {\bf r}_{4},\ t) \big>\
=\end{displaymath}
\begin{displaymath}\     {1\over {\sqrt2}}\psi_{+\ 1}({\bf
r}_{1},\ t)\cdot\psi_{+\ 2}({\bf r}_{2},\ t)\cdot\psi_{-\ 3}({\bf r}_{3},\ t)\cdot\psi_{-\ 4}({\bf
r}_{4},\ t)\cdot\big| +
\big>_{\rm 1}
\otimes \big| +\big>_{\rm 2}\otimes \big| -\big>_{\rm 3}\otimes \big| -\big>_{\rm
4}\end{displaymath}
\begin{equation}- \
{1\over {\sqrt2}}\psi_{-\ 1}({\bf r}_{1},\ t)\cdot\psi_{-\ 2}({\bf r}_{2},\ t)\cdot\psi_{+\
3}({\bf r}_{3},\ t)\cdot\psi_{+\ 4}({\bf r}_{4},\ t)\cdot \big| - \big>_{\rm 1} \otimes \big|
-\big>_{\rm 2} \otimes \big| +\big>_{\rm 3}\otimes \big| +\big>_{\rm 4},   \end{equation} The
particles are sent along four different coplanar directions and $\big|
+\big>_{\rm i}$ ($\big| -\big>_{\rm i}$) represents a spin up (down) state along the direction of
propagation of the
$i$th particle. Four ``product'' observables, each of which is associated to simultaneous spin
measurements performed on the four particles, are considered: $\Sigma^1_x\cdot \Sigma^2_x\cdot
\Sigma^3_x\cdot \Sigma^4_x$,
$\Sigma^1_y\cdot \Sigma^2_x\cdot
\Sigma^3_y\cdot \Sigma^4_x$,
$\Sigma^1_y\cdot \Sigma^2_x\cdot
\Sigma^3_x\cdot \Sigma^4_y$ and $\Sigma^1_x\cdot \Sigma^2_x\cdot
\Sigma^3_y\cdot \Sigma^4_y$, where
$\Sigma^i_x$ represents a Stern-Gerlach mesurement performed on the $i$th particle with the magnet
parallel to the
$X$ direction which is orthogonal to the plane of propagation of the particles and $\Sigma^i_y$
represents a Stern-Gerlach mesurement performed on the $i$th particle with the magnet parallel to
the
$Y$ direction which is orthogonal to the direction of propagation of the $i$th particle and to
$X$. 
$\big|
\Psi({\bf r}_{1},\ {\bf r}_{2},\ {\bf r}_{3},\ {\bf r}_{4},\ t)
\big>$ is simultaneous eigenstate of the four product observables for the eigenvalues $-1,\ -1,\
-1,\ +1$. As before, this imposes severe constraints on the results of local spin measurements. By
performing an EPR like reasoning similar to the previous ones, the assumption of locality implies
that these results must be determined before the measurements take place and may not
depend on which measurement is performed in distant regions on the other particles. Let us denote
$S_x^i$ and
$S_y^i$ the prediction assigned to the measurement of $\Sigma_x^i$ and $\Sigma_y^i$; the
prediction +1 corresponds to an up deflection inside the Stern-Gerlach device, and -1 to a
down deflection. These constraints are expressed as follows:
\begin{displaymath}S_x^1\cdot S_x^2\cdot S_x^3\cdot S_x^4\ =\ -1\end{displaymath}
\begin{displaymath}S_y^1\cdot S_x^2\cdot S_y^3\cdot S_x^4\ =\ -1\end{displaymath}
\begin{displaymath}S_y^1\cdot S_x^2\cdot S_x^3\cdot S_y^4\ =\
-1\end{displaymath}
\begin{equation}S_x^1\cdot S_x^2\cdot S_y^3\cdot S_y^4\ =\ +1.\end{equation}
The product of the three last equations gives:

\begin{equation}S_x^1\cdot S_x^2\cdot S_x^3\cdot S_x^4\ =\ +1,\end{equation} which together with
the first equation leads to a contradiction of the type +1=-1. The analysis of Bohm's trajectories
is similar to the two and three particle cases so that we shall not repeat it entirely here. For
instance, during the measurement of $\Sigma^1_x\cdot \Sigma^2_x\cdot
\Sigma^3_x\cdot \Sigma^4_x$ we get that the velocity of the first particle obeys the following
equation: 
\begin{displaymath} {dx_1\over dt }\ =\
{k^2\over 1\ +\ k^2t^2}\,tx_1\ + \end{displaymath}
\begin{displaymath} \{ \,-\,sh({\beta t^2 \over 1\
+\ k^2t^2} (-x_1\,-x_2\,-\,x_3\,+\,x_4))\,-\,sh({\beta t^2 \over 1\ +\
k^2t^2}(-x_1\,+\,x_2\,+\,x_3\,+\,x_4))\end{displaymath}
\begin{displaymath}\,+\,sh({\beta t^2 \over 1\
+\ k^2t^2} (+x_1\,-\,x_2\,+\,x_3\,+\,x_4))\,+\,sh({\beta t^2 \over 1\
+\ k^2t^2} (+x_1\,+\,x_2\,-\,x_3\,+\,x_4))\}\ /\end{displaymath}\begin{displaymath}
\{\,+\,ch({\beta t^2 \over 1\ +\ k^2t^2} (-x_1\,-x_2\,-\,x_3\,+\,x_4))\,+\,ch({\beta t^2
\over 1\ +\ k^2t^2} (-x_1\,+\,x_2\,+\,x_3\,+\,x_4))\end{displaymath}
\begin{displaymath}\,+\,ch({\beta t^2 \over 1\
+\ k^2t^2} (+x_1\,-\,x_2\,+\,x_3\,+\,x_4))\,+\,ch({\beta t^2 \over 1\
+\ k^2t^2} (+x_1\,+\,x_2\,-\,x_3\,+\,x_4))\}
\end{displaymath}\begin{equation}\cdot\{ {-k^2\over 1\
+\ k^2t^2}\,\alpha t^3\,+\,2\alpha t\}.\end{equation}
Similar equations are associated to the other velocities and to the other measurements. Only the
factor that contains exponential terms changes from equation to equation. It is easy to guess the
form of this factor because it obeys the following simple rules. To each outcome corresponds at the
numerator and at the denominator as well a product of exponential factors of the form
$exp(({\beta t^2
\over 1\ +\ k^2t^2} {(+/-x_1\,+/-\,x_2\,+/-\,x_3\,+/-\,x_4)\over 2}))$ where $x_i$ represents the
projection on the axis parallel to the local $i$th magnet. The sign of $x_i$ in the
exponent represents the spin value asymptotically reached during the local measurement. The weight
of this product is equal at the numerator and at the denominator as well to the probability to find
the corresponding outcome. When the sign of $x_i$ in the
exponent is negative, the whole product will also be negatively weighted at the numerator when we
consider the equation of evolution of
$x_i$. Otherwise, and also at the denominator, the products will have a positive weight.  This
can be checked directly in the previous equation and is also true for the two and three particles
cases.

As in the three particles case treated in the previous section, the integration of the velocities
leads to a paradoxical situation for any value of the initial position (excepted for some positions
that belong to a region of negligible weight). 

\section{Conclusions.}In conclusion, we showed in this paper how the elucidation of several
paradoxical situations related to the question of non-locality can be realised in the framework of
Bohm's interpretation and is finally reduced to the study of the topology of attractor basins of
the Bohmian dynamics. One could object that this explanation is obvious being given that no
satisfying relativistically covariant formulation of the Bohm's interpretation exists for
situations in which more than one (entangled) particles are involved. This criticism is valid: in
the present approach, we systematically considered the Schr\"odinger equation, in which a unique
time appears, similar to the Newtonian, absolute, time. Now, a theorem due to Hardy \cite{hardy2}
shows the impossibility to build a Lorentz-Invariant realistic theory that would mimick the
predictions of quantum mechanics. In order to preserve realism it is therefore necessary to
reintroduce a kind of absolute or etheric time, attached to an absolute or special frame of
reference, which answers to the previous criticism. The interesting question is then, could the
Bohmian interpretation help us to conceive an experiment that would reveal the existence of this
quantum ether. The original Bohmian interpretation is too ad hoc to allow for such a possibility,
but slightly modified versions of the interpretation (in which the randomisation of the hidden
variable, here the position of the particle, according to the $\psi^2$ distribution is not
guaranteed for all times) would allow us to conceive such experiments \cite{durt2}. A realistic
interpretation {\it \`a  la Bohm} would then open the door to the conception of a quantum version of
the Michelson-Morley experiment.

\section{Appendix: Bohm's solution for the passage of one spin 1/2 particle through a
Stern-Gerlach device.}  Let us describe the wave-function of a spin 1/2 particle as a
superposition of a spin up and a spin down components along the direction $Z$ of the magnetic field
of the Stern-Gerlach device:
$\Psi({\bf r},\ t)\,=\,\psi_+({\bf r},\ t)\,\big| +\big>\,+\,\psi_-({\bf r},\ t)\,\big|-\big>$.
 The Pauli-Schr\"odinger equation describes the evolution of this wave-function in the presence of
an external magnetic field ${\bf B}$: 
\begin{equation} i\hbar\, \partial_t\,\Psi({\bf r},\ t)\,=\, -{\hbar^2 \over 2m}
\Delta\,\Psi({\bf r},\ t)\, -\, \mu\, {\bf B}\cdot{\bf  \Sigma}\,
\Psi({\bf r},\ t),\end{equation}
where ${\bf \Sigma}$ represents the Pauli matrices and $\mu$ is the gyromagnetic coupling constant
of the (neutral) particle. In a Stern-Gerlach device, the field is always parallel to the
$Z$ direction, and its gradient is constant ($B_x\,=\,B_y\,=\,0$, and $B_z\,=\,(a_0\,+\,a_1\,z)$.
Then, the evolutions of the two spin components decouple and we get that:
 \begin{equation} i\hbar\, \partial_t\,\psi_{+/-}({\bf
r},\ t)\,=\, -{\hbar^2 \over 2m} \Delta\,\psi_{+/-}({\bf r},\ t)\, -/+\, \mu\, 
(a_0\,+\,a_1\,z)\psi_{+/-}({\bf r},\ t)\end{equation}
When the incoming wave-packet is initially Gaussian shaped, this equation is separable in Cartesian
coordinates, and we find :
\begin{displaymath} i\hbar\, \partial_t\,\psi^x_{+/-}( x,\ t)\,=\,
-{\hbar^2 \over 2m} \partial_x^2\,\psi_{+/-}^x( x,\ t)\end{displaymath}
\begin{displaymath} i\hbar\, \partial_t\,\psi^y_{+/-}( y,\ t)\,=\,
-{\hbar^2 \over 2m} \partial_y^2\,\psi_{+/-}^y( y,\ t)\end{displaymath}
\begin{equation} i\hbar\, \partial_t\,\psi_{+/-}^z( z,\ t)\,=\,
-{\hbar^2 \over 2m} \partial_z^2\,\psi^z_{+/-}( z,\ t)\, -/+\, \mu\, 
(a_0\,+\,a_1\,z)\psi^z_{+/-}( z,\ t)\end{equation}
The two first equations correspond to a free propagation, and in the case of an
initial Gaussian wave packet, their solution is well known and given in many standard textbooks
of quantum mechanics, sothat we shall not discuss how we obtain their contribution to the
solution. The third equation is less common. We shall show, following Bohm himself \cite{bohm1} how
to solve it. Let us try to find a generalised plane wave solution of the form $\psi_{+/-}^z( z,\ t)\
=\ f_{+/-}( z,\ t)\cdot exp\,i(kz\,-\,\hbar k^2 t /2m)$. Then, $f$ fulfills :
\begin{equation}i\hbar\, \partial_t\,f_{+/-}^z( z,\ t)\,=\,
-{\hbar^2 \over 2m} (\partial_z^2\,+/-\,2ik\partial_z)f_{+/-}^z( z,\ t)\, -/+\,
\mu\,  (a_0\,+\,a_1\,z)f^z_{+/-}( z,\ t)\end{equation} If we could neglect the
derivatives, we would have $f_{+/-}^z( z,\ t)\,=\,exp(+/-i{\mu\,t\over\hbar}
(a_0\,+\,a_1\,z))$, but then the derivatives are not zero but yeld ${\hbar^2 \over
2m}\{+/- kt{2a_1\mu\over \hbar}\,+\,({a_1\mu t\over\hbar})^2\}$. This term
depends on the time only so that we can compensate it by multiplying the
postulated value of $f$ by $exp(i\{-/+ {a_1\mu\over 2m}kt^2\,-\,{\mu^2 a_1^2
\over 6m\hbar}t^3\})$. We find so the exact solution :
\begin{equation}\psi_{+/-}^z( z,\ t)\ =\ exp(i(kz\,-\,\hbar k^2 t
/2m\ +/-\ {\mu\,t\over\hbar}
(a_0\,+\,a_1\,z)\ -/+ {a_1\mu\over 2m}kt^2\,-\,{\mu^2 a_1^2
\over 6m\hbar}t^3)\end{equation}
When $t\,=\,0$, this wave is plane: $\psi_{+/-}^z( z,\ 0)\,=\,exp(i\,kz)$.
Thanks to the linearity of the Schr\"odinger equation, the general solution is a
superposition of these generalised plane waves with as weight the Fourier
components at time zero. The initial packet being Gaussian shaped, we have :
\begin{displaymath} \psi^z_{+/-}( z,\ t\,=\,0)\ =\ \sqrt{{1\over
(2\sqrt{\pi}\,\delta r_0)}}\ exp({-z^{2}\over 4\delta r_0^2})\end{displaymath} 
Its Fourier transform is easily obtained thanks to the well-known formula :
\begin{displaymath}\int_{v\,\in\,{\bf R}}
\,dv\,exp(-av^2\,+\,bv)\,=\,\sqrt{{\pi\over a}}exp({b^2\over
4a})\end{displaymath} It gives $\sqrt{{\delta r_0\over
\sqrt{\pi}}}\ exp(- \delta r_0^2\,k^{2})$

The wave function at time $t$ is thus :
\begin{displaymath}\psi_{+/-}^z( z,\ t)\ =\ {1\over \sqrt{2\pi}}\int
\,dk\,exp(i(kz\,-\,\hbar k^2 t/2m\ +/-\ {\mu\,t\over\hbar} (a_0\,+\,a_1\,z)\ -/+
{a_1\mu\over 2m}kt^2\,-\,{\mu^2 a_1^2 \over
6m\hbar}t^3)\end{displaymath}\begin{equation}\times\sqrt{{\delta r_0\over
\sqrt{\pi}}}\ exp(- \delta r_0^2\,k^{2})\end{equation} We can let go out of the
integral the terms independant of $k$, and reorder the other ones:
\begin{displaymath}\psi_{+/-}^z( z,\ t)\ =\ {\sqrt{2\delta r_0}\over
2\sqrt{\pi^{3/2}}}\ exp(+/-{\mu\,t\over\hbar}
(a_0\,+\,a_1\,z)\, -\,{\mu^2 a_1^2
\over 6m\hbar}t^3)\end{displaymath}\begin{equation}\times
\int \,dk\,exp(-k^{2}( \delta r_0^2\,+\,i\hbar\, t/2m)\,+\,ik(z\,-/+
{a_1\mu\over 2m}t^2) )\end{equation}

We get :
\begin{displaymath}\psi_{+/-}^z( z,\ t)\ =\ \sqrt{1\over
2\sqrt{\pi}(\delta r_0\,+\,i{\hbar\, t\over 2m\delta r_0})}\ exp(+/-{\mu\,t\over\hbar}
(a_0\,+\,a_1\,z)\, -\,{\mu^2 a_1^2
\over 6m\hbar}t^3)\end{displaymath}
\begin{equation}\times exp-{(z\,-/+ {a_1\mu\over 2m}t^2)^2 \over 4( \delta
r_0^2\,+\,i\hbar\, t/2m)}\end{equation} 
Using the equalities 
\begin{displaymath} {1\over \delta r_0^2\,+\,i\hbar\, t/2m}\,=\, {(\delta
r_0^2\,-\,i\hbar\, t/2m)\over (\delta r_0^4\,+\,\hbar^2\,
t^2/4m^2)}\end{displaymath}
\begin{displaymath}\sqrt{1\over 2\sqrt{\pi}(\delta r_0\,+\,i{\hbar\, t\over 2m\delta r_0})}\ =\
\sqrt{{1\over 2\sqrt{\pi}\sqrt{(\delta r^2_0\,+\,{\hbar^2\, t^2\over 4m^2\delta
r^2_0})}}\,\cdot\,  exp\,-i\ arccos {\delta r_0\over \sqrt {(\delta
r^2_0\,+\,{\hbar^2\, t^2\over 4m^2\delta r^2_0})}}}\end{displaymath} and 
\begin{displaymath} \delta r^2_t\ =\ \delta r^2_0\cdot(1\,+\,{\hbar^2
t^2\over 4m^2\delta r^4_0 })\end{displaymath}
 we get the contribution of 
$\psi_{+/-}^z( z,\ t)$ to the wave-function: 
\begin{displaymath}\psi_{+/-}^z( z,\ t)\ =\ {1\over
(2\sqrt{\pi}\,\delta r_t)^{1/2}}\cdot  exp-{(z\ -/+\ a_1\mu t^2/2m)^2\over 4\delta^2
r_t}\end{displaymath}\begin{equation}\times exp\ i\,({\hbar t\ (z\ -/+\ a_1\mu
t^2/2m)^2 \over 2m\delta^2r_0\  4\delta r^2_t}\ -\ {arccos{\delta r_0\over\delta
r_t}\over 2}\ +/-\ {\mu t\over \hbar}\cdot(a_0\ +\ a_1 z)-\,{\mu^2 a_1^2
\over 6m\hbar}t^3)\end{equation} Notice that $\psi_{+/-}^y( y,\ t)$ is
obtained in exactly the same way, with the requirement that $a_0\,=\,a_1\,=\,0$.

In conclusion,  
\begin{displaymath}\psi_{+/-}({\bf r},\,t)\ =\ {1\over (2\sqrt{\pi}\,\delta r_t)^{3/2}}\
exp-{(x\,-\,v_0t)^2\over 4\delta r^2_t}\end{displaymath}\begin{displaymath}
\times\ exp\,i({\hbar t\ (x\,-\,v_0t)^2 \over 2m\delta^2r_0\ 4\delta r^2_t}\, +\,
k_0 x\ -\,{\hbar k^2_0t\over 2m}\ -\ {arccos{\delta r_0\over\delta r_t}\over 2})
\end{displaymath}  \begin{displaymath}\times\ \ exp-{y^2\over 4\delta
r^2_t}\cdot exp\,i({\hbar t\ y^2 \over 2m\delta^2r_0\ 4\delta r^2_t}\ -\
{arccos{\delta r_0\over\delta r_t}\over 2}) \end{displaymath} 
\begin{displaymath}\times exp-{(z\ -/+\ a_1\mu t^2/2m)^2\over 4\delta^2
r_t}\cdot exp\,i({\hbar t\ (z\ -/+\ a_1\mu t^2/2m)^2 \over 2m\delta^2r_0\ 
4\delta r^2_t}\ -\ {arccos{\delta r_0\over\delta r_t}\over
2})\end{displaymath}\begin{equation}\times exp( +/-\ {\mu t\over \hbar}\cdot(a_0\
+\ a_1 z)\,-\,{\mu^2 a_1^2 \over 6m\hbar}t^3) \end{equation}
with $ \delta r^2_t\ =\ \delta r^2_0\cdot(1\,+\,{\hbar^2
t^2\over 4m^2\delta r^4_0 })$. The interpretation of this solution which appeared
originally in \cite{bohm1}) is straightforward: the wave-packet of the particles of spin up
(down) diffuses under the influence of the free part of the Schr\"odinger evolution (the Laplacian
term) and is simultaneously uniformly accelerated upwards (downwards). It ``feels'' effectively a
potential due to the gyromagnetic coupling which varies linearly in $z$, and changes of sign with
the
$z$-spin. 

After a time $\tau$, the particles leave the Stern-Gerlach device, and the Gaussian
packets move freely, diffusing around centers which conserve the velocity that
they possessed when leaving the magnets. The dependence in $z$ is thus, up to
irrelevant global phases:

\begin{displaymath}   exp-{(z\ -/+\ a_1\mu \tau t/m\ +/-\ a_1\mu \tau^2/2m)^2\over
4\delta r^2_t}\cdot\end{displaymath} \begin{equation} exp\,i({\hbar t\ (z\ -/+\
a_1\mu \tau t/m\ +/-\ a_1\mu \tau^2/2m)^2 \over 2m\delta^2r_0\  4\delta r^2_t}\
+/-\,a_1\mu \tau /m\,(z\,-/+\,a_1\mu \tau^2/2m))                   
\end{equation}while the dependence in $x$ and $y$ is the same as before.

In a typical single Stern-Gerlach experiment involving
silver atoms, \begin{itemize}
\item the mass of the incoming atom is equal to 1,80 $10^{-22}$
grams
\item the gyromagnetic coupling constant $\mu$ is equal to 9,27 $10^{-21}$ gram
cm$^{2}$ sec$^{-2}$ Gauss$^{-1}$
\item the gradient of the magnetic field along the $Z$ axis is equal to $10^{4}$ Gauss
cm$^{-1}$
\item the length of the magnet is equal to 10 cm
\item the velocity of the incoming particle along the $X$ axis is equal to $10^{4}$ cm
sec$^{-1}$ 
\item the time of flight through the magnet $\tau$ is equal to 10$^{-3}$ sec
\item $\hbar$ is equal to 1,05 10$^{-27}$ ergsec
\item the spreading of the incoming particle $\delta r_0$ is equal to 10$^{-3}$ cm.
\end{itemize}
Then, \begin{equation}k\ =\ {\hbar\over 2m\delta^2r_0}\ =\
2,91\,{\rm sec}^{-1}\end{equation}
\begin{equation}\alpha\ =\ a_1\mu/2m\ =\ 2,58\,10^5 {\rm cm\
sec}^{-2}\end{equation} 
\begin{equation}\beta\ =\ a_1\mu/m\delta r^2_0\ =\ 5,15\,10^{11} {\rm
cm}^{-1}\,{\rm sec}^{-2}\end{equation}
The heights of the centers of the outgoing bundles moving up and downwards
after the time $\tau$ are equal to $+/-\, \alpha\,\tau^2\ =\ +/-\,0,258$ cm. Their
$Z$-velocities are $+/-\,2\alpha\,\tau\ =\ +/-\,5,15\,$m sec$^{-1}$. The
spreading is then, taking the diffusion into account, $\delta
r_0\cdot\sqrt{1\,+\,k^2\tau^2}\  =\  10^{-3}\cdot\sqrt{1\,+\,(2,91\,10^{-3})^2}$
cm (approximatively 10$^{-3}$ cm),which is very small in comparison to the
distance between the bundles (0,515 cm), sothat these bundles are clearly
separated. If we place a screen at a distance 1 meter after the magnet, it is
reached after 10$^{-2}$ seconds, and the spots are separated by a distance of  
 order of 11 cm, while the extent of a single spot is still of the order of 
10$^{-3}$ cm.

\section{Appendix 2: The two particles case.}

The Pauli-Schr\"odinger equation allows us to describe the evolution of the
wave-function inside the magnets of the devices:
\begin{equation}i \hbar\, \partial_t\, \Psi({\bf r}_{L},\, {\bf r}_{R},\, t) =
-{\hbar^2 \over 2m} (\Delta_{L}\, +\,  \Delta_{R})\Psi({\bf r}_{L},\, {\bf
r}_{R},\, t)\, +\, (\mu\, {\bf B}_{L}\cdot{\bf  \Sigma}_{L}\, +\, \mu\,{\bf
B}_{R}\cdot{\bf \Sigma}_{R}) \Psi({\bf r}_{L},\, {\bf r}_{R},\, t) \end{equation}
with \begin{displaymath}\Delta_{L}(\Delta_{R})(\psi_{+/-\ L}({\bf r}_{L},\
t)\cdot\psi_ {+'/-'\ R}({\bf r}_{R},\ t)\cdot\big| +/-
\big>_{\rm L} \otimes \big| +'/-'\big>_{\rm R})\
=\end{displaymath}\begin{displaymath}\  (\Delta_{L}\psi_{+/-\ L}({\bf r}_{L},\
t))\cdot\psi_{+'/-'\ R}({\bf r}_{R},\ t)\cdot\big| +/- \big>_{\rm L} \otimes \big|
+'/-'\big>_{\rm R} \end{displaymath}
\begin{equation}(\psi_{+/-\ L}({\bf r}_{L},\ t)\cdot(\Delta_{R}\psi_{+'/-'\
R}({\bf r}_{R},\ t))\cdot\big| +/- \big>_{\rm L} \otimes \big| +'/-'\big>_{\rm R})
\end{equation} 

where $\Delta_{L/R}$ is the Laplacian operator in the
Left/Right coordinates, ${\bf B_{L/R}}$ is the magnetic field in the
Left/Right regions, while the components of ${\bf \Sigma_{L/R}}$  are the
Sigma matrices of Pauli. When the magnetic fields ares parallel
 to the $Z$ axes, only the third Sigma matrices appear, which are defined by: 
\begin{displaymath} {\bf \Sigma}^3_{L(R)}\psi_{+/-\ L}({\bf r}_{L},\
t)\cdot\psi_{+'/-'\ R}({\bf r}_{R},\ t)\cdot\big| +/- \big>_{\rm L} \otimes
\big| +'/-'\big>_{\rm R}\ =\end{displaymath} \begin{equation}\
+/-\,(+'/-')\psi_{+/-\ L}({\bf r}_{L},\ t)) \cdot\psi_{+'/-'\ R}({\bf r}_{R},\
t)\cdot\big| +/- \big>_{\rm L} \otimes \big| +'/-'\big>_{\rm R}\end{equation}
Then, it is easy to check that the Pauli-Schr\"odinger equation is separable into the six spatial
coordinates and that the general solution, when the wave-packet is initially Gaussian, can be
expressed in terms of the solutions associated to the single particle case that we obtained in the
previous appendix.   

Let us now derive the conservation equation associated to this evolution law. We can write it in
the form:
\begin{equation}i \hbar\, \partial_t\, \Psi({\bf r},\, t) =
-{\hbar^2 \over 2m} \Delta\Psi({\bf r},\, t)\, +\, 
\mu\, {\bf B}\cdot{\bf  \Sigma}\Psi({\bf r},\, t) \end{equation}
where the Laplacian is a 6-dimensional Laplacian, the magnetic 
field a 6-dimensional field. It is worth noting that ${\bf  \Sigma}$ is a self-adjoint operator on
the spin space (isomorph to ${\bf C}^4$). The adjoint
(transposate-conjugate) of the previous equation is:
\begin{equation}-i \hbar\, \partial_t\, \Psi^{\dagger}({\bf r},\, t) =
-{\hbar^2 \over 2m} \Delta\Psi^{\dagger}({\bf r},\, t)\, +\, 
\Psi^{\dagger}({\bf r},\, t)\,\mu\, {\bf B}\cdot{\bf  \Sigma} \end{equation}
where we used the fact that ${\bf  \Sigma}\ =\ {\bf  \Sigma}^{\dagger}$.
Let us now multiply the equation of evolution by $\Psi^{\dagger}$, its adjoint equation
by $\Psi$ and take their difference, we get:
\begin{equation}i \hbar\, \partial_t\,\Psi^{\dagger}({\bf r},\, t)\cdot \Psi({\bf
r},\, t) = -{\hbar^2 \over 2m} \Psi^{\dagger}({\bf r},\, t)\cdot(\Delta\Psi({\bf
r},\, t))\, -\,(\Delta\Psi^{\dagger}({\bf r},\,t))\cdot\Psi({\bf r},\, t) 
 \end{equation} 

We can rewrite the second member of the last equation as the 6 dimensional
divergency of a 6 dimensional flux:
\begin{displaymath}\Psi^{\dagger}({\bf r},\, t)\cdot(\Delta\Psi({\bf r},\, t))\,
-\,(\Delta\Psi^{\dagger}({\bf r},\,t))\cdot\Psi({\bf r},\,
t)\,=\end{displaymath}\begin{equation}div(\Psi^{\dagger}({\bf r},\, t)\cdot{\rm
grad}\Psi({\bf r},\, t)\, -\,{\rm grad}\Psi^{\dagger}({\bf
r},\,t)\cdot\Psi({\bf r},\, t))\end{equation} Now, $\Psi^{\dagger}({\bf
r},\,t)\cdot{\rm grad}\Psi({\bf r},\, t)$ is the complex conjugate of ${\rm
grad}\Psi^{\dagger}({\bf r},\,t)\cdot\Psi({\bf r},\, t)$ so that we obtain the
equation of conservation $\partial_t\,\rho\ =\ div({\bf {\cal J}})$ where:
\begin{equation} \rho\,=\,\Psi^{\dagger}({\bf r},\, t)\cdot \Psi({\bf
r},\, t)\end{equation}or, in another notation:
\begin{displaymath}\rho({\bf r}_{L},\ {\bf r}_{R},\ t)\
=\ \big<\Psi({\bf r}_{L},\ {\bf r}_{R,\ t)}\big|
\Psi({\bf r}_{L},\ {\bf r}_{R,\ t)} \big>\
\end{displaymath}and
\begin{equation}{\bf {\cal J}}\,=\,(\hbar/m)\Im (\Psi^{\dagger}({\bf r},\,
t){\rm grad}\Psi({\bf r},\, t))\end{equation}where $\Im(z) $ the imaginary part of $z$. 

The three and four particles cases can be treated in a similar way.


\begin{thebibliography}{99}
 
\bibitem{bell} J.S. Bell: "On the EPR paradox", {\it Physics} {\bf 1} (1964) 195.

\bibitem{bohm1} D. Bohm: "Quantum theory", Prentice Hall,Englewood cliffs, N.J.
(1951).

\bibitem{bohm2} D. Bohm: "A suggested interpretation of quantum theory in terms of
hidden variables", {\it Physical Review} {\bf 85} (1952) 166.

\bibitem{clauser} J.F. Clauser and M.A. Horne: "Experimental consequences of
objective local theories", {\it Physical Review D} {\bf 10} (1974) 526.

\bibitem{dewdney1} C. Dewdney, P.R. Holland, A. Kyprianidis and J.P. Vigier:
"Spin and non-locality in quantum mechanics", {\it Nature} {\bf 336} n$^\circ$6199 (1988)
53¤. 
\bibitem{dewdney2} C. Dewdney: ``Constraints on quantum hidden-variables and the Bohm theory'',
{\it Journal of Physics A} {\bf 25} (1992) 3615.

 
\bibitem{durt} T.Durt: ``Three interpretations of the violation of Bell's inequalities'',
{\it Foundations of  Physics}, {\bf 27} (1997) 415.

\bibitem{durt2}  T. Durt: ``About the possibility of supraluminal transmission of information in the
Bohm-Bub theory'' {\it Helvetica Physica Acta}, {\bf 72} (1999) 356.  


\bibitem{epr} A. Einstein, B. Podolsky, and N. Rosen: ``Can quantum-mechanical description of
physical reality be considered complete?'', {\it Physical Review} {\bf 47} (1935) 777.

\bibitem{ghz}D.M. Greenberger, M.A. Horne, A. Shimony and A. Zeilinger: ``Bell's theorem without
inequalities'', {\it American Journal of Physics} {\bf 58} (12) (1990) 1131.

\bibitem{hardy}L. Hardy: ``A new way to obtain Bell's inequalities'', {\it Physics Letters A} {\bf
161} (1991) 21.

\bibitem{hardy2}L. Hardy: ``Quantum Mechanics, Local Realistic Theories, and Lorentz-Invariant
Realistic Theories'', {\it Physical Review Letters} {\bf 68} (1992) 20.


\bibitem{mermin}N. D. Mermin: ``What's wrong with these elements of reality?'' {\it Physics Today}
{\bf 43} (6) (1990) 9. 

\bibitem{peres} A. Peres: ``Incompatible results of quantum measurements'', {\it Physics Letters A}
{\bf 151} 107.

\bibitem{pito} I. Pitowsky: "Quantum probability. Quantum logic",
Springer Verlag, Lecture Notes in Physics (1989). 



\end{thebibliography}
 \end{document}